\documentclass{evn2002_mag}
\begin{document}
\title{Distant star forming galaxies, next
  generation radio telescopes and the radio universe before
  re-ionisation}

   \author{M.A. Garrett\inst{1}}

   \institute{Joint Institute for VLBI in Europe, Postbus 2, 7990~AA
   Dwingeloo, the Netherlands}

\titlerunning{Star Forming Galaxies and Next Generation radio telescopes}

\offprints{garrett@jive.nl} 

\abstract{I present the various capabilities of upgraded and next
  generation radio telescopes, in particular their ability to detect
  and image distant star forming galaxies. I demonstrate that e-MERLIN,
  EVLA and LOFAR can detect systems similar to Arp 220 out to
  cosmological distances. The SKA can detect such systems out to any
  reasonable redshift that they might be expected to exist. Employing
  very long integration times on the multiple-beam SKA will require the
  array to be extended beyond the current specification - simply to
  avoid confusion noise limitations at 1.4~GHz. Other arguements for
  extending the SKA baseline length are also presented. As well as
  going ``deeper'' all these instruments (especially LOFAR and the SKA)
  will also go ``wider'' - detecting many tens of thousands of galaxies
  in a single day's observing. I briefly comment on the prospects of
  detecting radio emission at much earlier epochs, just before the
  epoch of re-ionisation.}

   \maketitle
%

\section{Introduction}

The continuum sensitivity of radio astronomy instruments is set to
improve by at least an order of magnitude over the next 10 years.
Developments include the broad-banding of existing facilities (e.g.
e-MERLIN and the EVLA) or the design and construction of entirely new,
next generation instruments, such as the Low Frequency Array (LOFAR)
and the Square Km Array (SKA). VLBI arrays are also expected to take
advantage of the increasing capacity of disk-based recording systems,
and the real-time connection of antennas and correlators via commercial
optical fibre networks. These upgraded and next generation telescopes
will routinely reach noise levels which are at the very limit of what
is now feasible with existing facilities.  Deep surveys of the microJy
radio sky (e.g. Fomalont et al. 2002) suggest that the radio emission
will be associated with a dominant population of moderate and high
redshift galaxies that are subject to on-going, massive star formation.

In this paper I consider and contrast the capabilities of next
generation and upgraded radio telescopes, in particular their ability
to detect and image these faint and distant systems out to redshift 6.
I also briefly comment on the prospects of detecting radio sources with
the SKA before the epoch of re-ionisation.


\section{Technical and Scientific Assumptions} 

In this paper I use the following sources of information regarding
telescope parameters:
\begin{itemize}
\item VLA Expansion Project (see {\tt
  www.nrao.edu/evla}),
\item e-MERLIN Science Case \& Technical
  Specification (see {\tt www.merlin.ac.uk/e-merlin/}),
\item LOFAR Basic
Specifications ({\tt www.lofar.org}), 
\item SKA Technical Specification \\ ({\tt
  www.skatelescope.org}), 
\item $e$EVN ({\tt
  www.evlbi.org/eEVN/sens\_uvcov/eEVN.htm}).
\end{itemize} 

I make use of two spectral energy distributions (SED) that I have
constructed from publicly available data of two well known, nearby star
forming galaxies - Arp 220 and M82. While the radio SEDs of these two
galaxies are similar, they differ in detail {\it e.g.} the
low-frequency radio spectrum of Arp 220 is much flatter than M82. It is
not yet clear which (if either) is more representative of the
population of high-z star forming galaxies. It is certainly worth
remembering that the most distant systems have star formation rates
(SFR) one or two orders of magnitude greater than Arp 220 and M82
respectively.

I assume and extrapolate the 1.4~GHz source counts of Richards (2000),
imposing a simple frequency dependence on the counts in order to
investigate LOFAR source counts at 200~MHz. Throughout this paper I
assume the currently ``preferred'' cosmological model ($\Omega_{{\rm
    m}}=0.3$, $\Omega_{\Lambda}=0.7$, $H_{{\rm 0}}=70$~km/sec/Mpc).

   \begin{figure}[h]
   \centering
   \vspace{6.5cm}
   \includegraphics{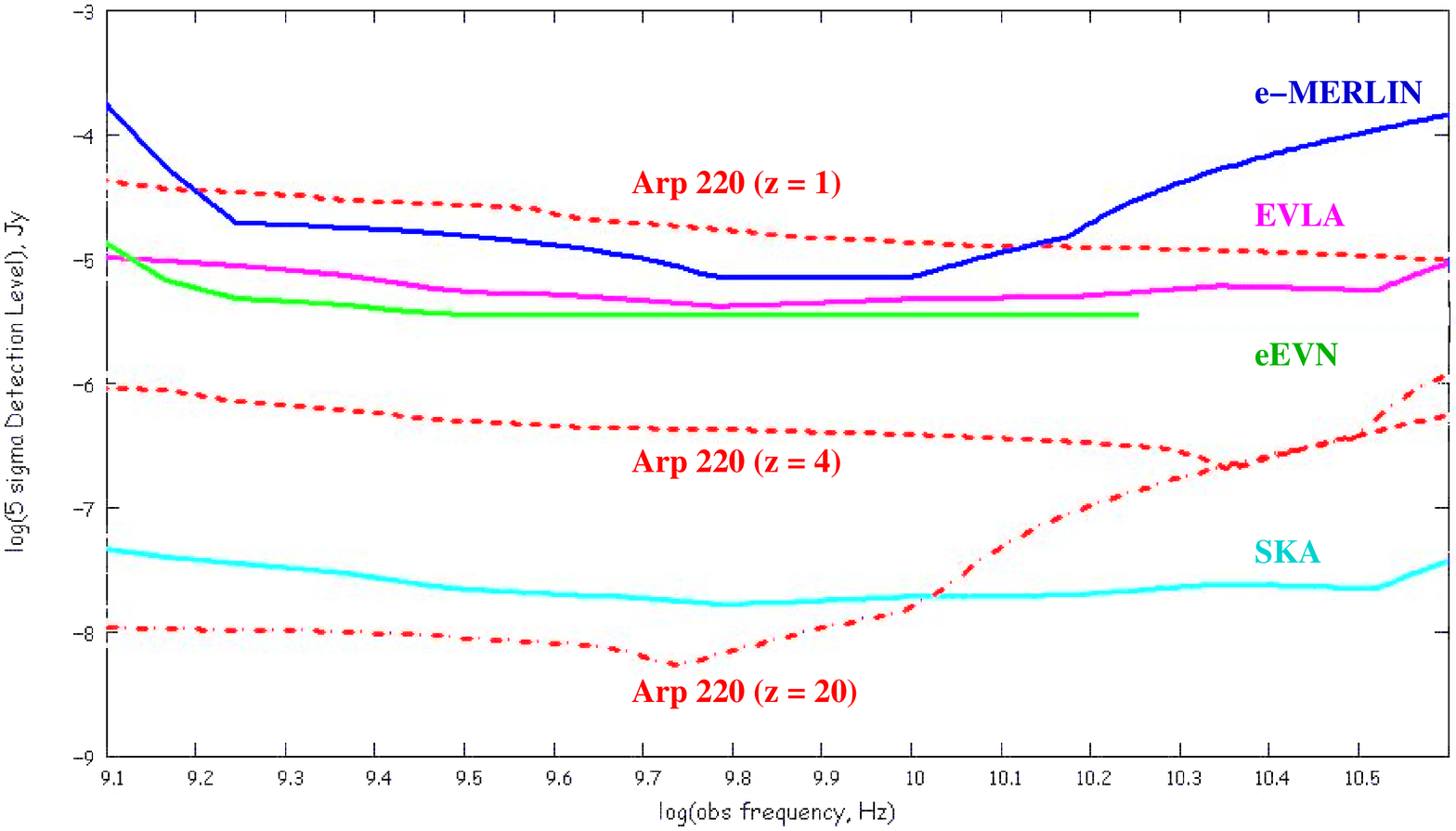}
      \caption{The $5\sigma$ detection limit (solid lines) plotted for e-MERLIN, the
        EVLA, $e$EVN and SKA as a function of observing frequency.
        Superimposed the SED of Arp 220 (dotted/dashed lines) projected
        to $z \sim 1$, $z \sim 4$ and $z \sim 20$. }
\label{sens_z}
   \end{figure}
%

\section{Detection of star forming galaxies at cm wavelengths} 

\subsection{Continuum sensitivity and Star Formation Rates} 

Fig.~\ref{sens_z} shows the 12 hour ($5\sigma$) sensitivity of
e-MERLIN, the EVLA, the $e$EVN and SKA at cm wavelengths as a function
of observing frequency. Superimposed on top of this is the SED of Arp
220, projected to redshifts 1, 4 and 20. The plot shows that the
upgraded instruments can in principle detect Arp 220 (SFR~$ \sim
250$~M$_{\odot}$/yr) out to $z \sim 1$ or beyond in only 12 hours. The
performance of the $e$EVN is particularly encouraging; with the current
high sensitivity array augmented by the 64-m Sardinia Radio Telescope,
the upgraded Lovell 76-m telescope and the 45-m Yebes telescope, the
$e$EVN out-performs the EVLA. Naturally some of these faint radio
sources will be resolved-out by the higher resolution $e$EVN (but see
section~\ref{size}). The SKA goes about 2 orders of magnitude deeper
than any of the other arrays, easily detecting Arp 220 at $z \sim 6$ in
12 hours. Note also that at high-z the peak in the FIR (rest-frame)
begins (in principle) to redshift into the radio part of the spectrum
(but see section \ref{caveats} for some caveats).  Fig.~\ref{sens_z} also
suggests that deep radio surveys will be conducted over a broad range
of frequency space, lower frequencies may take advantage of the
possible steep spectrum nature of the sources (especially at high-z,
see section~\ref{caveats}) but higher frequencies benefit from the
availability of large continuum bandwidths.

   \begin{figure}[h]
   \centering
   \vspace{6cm}
   \includegraphics{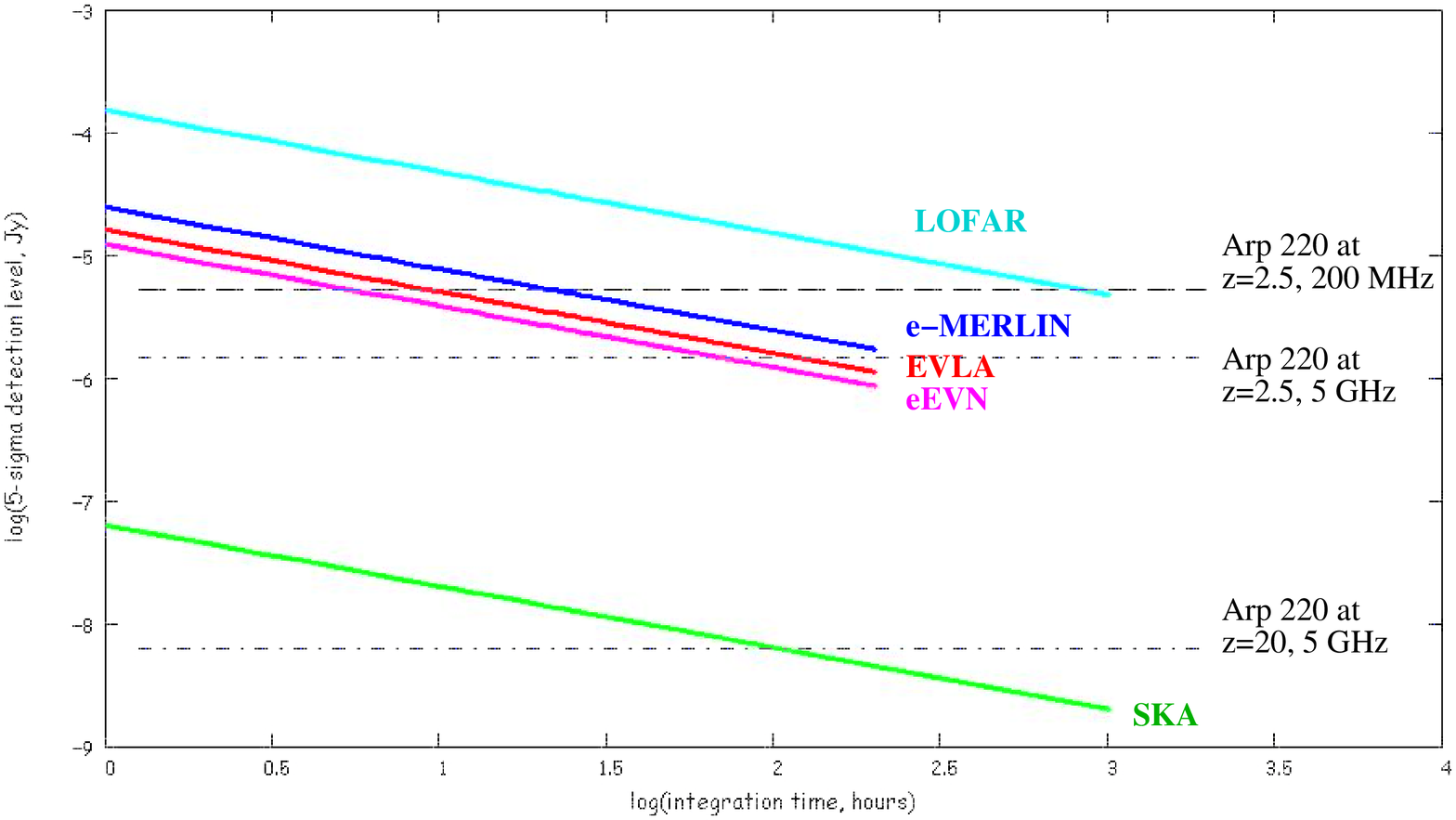}
      \caption{The ($5\sigma$) sensitivity (solid lines) of various
   instruments as a function of integration time. The flux density of
   Arp 220 (dotted/dashed lines) is indicated for three
   redshift/frequency combinations.}
\label{sens_dt}
   \end{figure}
%

\subsection{Multiple beams and long integration times} 

So far we have assumed {\it typical} integration times of $\sim
12$~hours. However, this is really a lower limit for deep field
surveys, especially for next generation telescopes such as LOFAR and
SKA. These instruments will possess a multiple beam capability,
permitting simultaneous, full-sensitivity observations to be made over
widely separated regions of the sky. It is expected that some beams
will be dedicated to particular areas of research, permitting very long
integration times to be employed for particular deep field surveys.

   \begin{figure}[b]
   \centering
   \vspace{6.5cm}
   \includegraphics{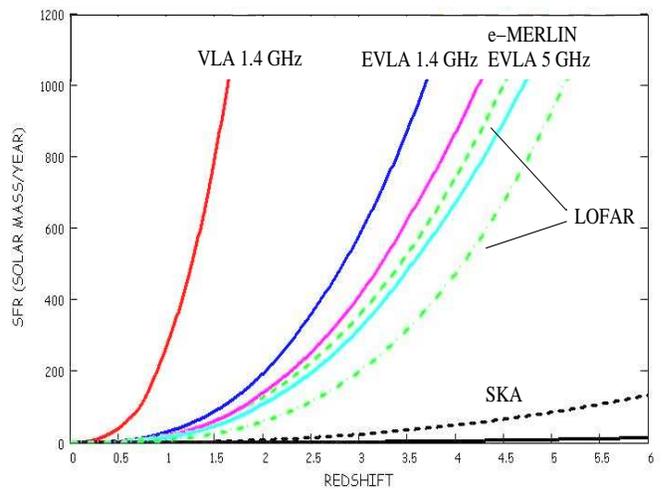}
      \caption{A plot of SFR versus redshift assuming in all but one
        case an Arp 220 SED. Various instruments are included - the
        integration times, observing frequencies are listed within the
        brackets (from left to right): VLA (1.4 GHz, 2 days), the EVLA
        at (1.4 GH, 2 days), EVLA (5 GHz, 2 days), LOFAR (200 MHz, 1000
        hours), e-MERLIN (5 GHz, 18 days, Arp 220), LOFAR (200 MHz,
        1000 hours, $\alpha \sim 0.8$ spectral index), SKA (5 GHz, 12
        hours), SKA (5 GHz, 1000 hours). }

\label{sfr}
   \end{figure}
%
   
   Fig.~\ref{sens_dt} shows the ($5\sigma$) sensitivity of various
   instruments as a function of integration time. Without a multiple
   beam capability, I have limited the longest EVLA and e-MERLIN
   observations to 100 hours.  Fig.~\ref{sens_dt} demonstrates that the
   sensitivity of LOFAR is comparable to e-MERLIN and the EVLA,
   provided very long integration times (up to 1000 hours) can be
   successfully employed.  LOFAR deep field surveys are optimised (in
   terms of survey depth) at the high-end of the LOFAR band, $\sim
   200$~MHz.  In principle, the SKA appears to be capable of detecting
   star forming galaxies such as Arp 220 out to any reasonable redshift
   that they might be expected to exist (but again see
   section~\ref{caveats}).
   
   Fig.~\ref{sfr} presents the SFR that can be probed by various
   instruments assuming the Condon (1992) relation between SFR and
   radio luminosity. In this plot we present the current sensitivity of
   the VLA and WSRT at 1.4~GHz (assuming typical integration times of a
   few days). Only SFR in excess of $1000$~M$_{\odot}$/yr are currently
   detectable beyond redshift $\sim 1.6$. The SFR probed by the deepest
   LOFAR and e-MERLIN/EVLA observations (the latter being conducted at
   5~GHz) are similar, the exact details depending on the radio SED of
   the sources. Plots are presented for LOFAR assuming the (flattish)
   SED of Arp 220, and a steeper ($\alpha\sim -0.8$) spectral index.
   The SKA is able to detect normal star forming galaxies with SFR~$ \sim
   1$~M$_{\odot}$/yr (comparable to the Milky Way) out to redshift 2
   (assuming integration times of 1000 hours).

\subsection{The number of detectable sources in the FoV}  

\begin{figure}[h]
   \centering
   \vspace{6cm}
   \includegraphics{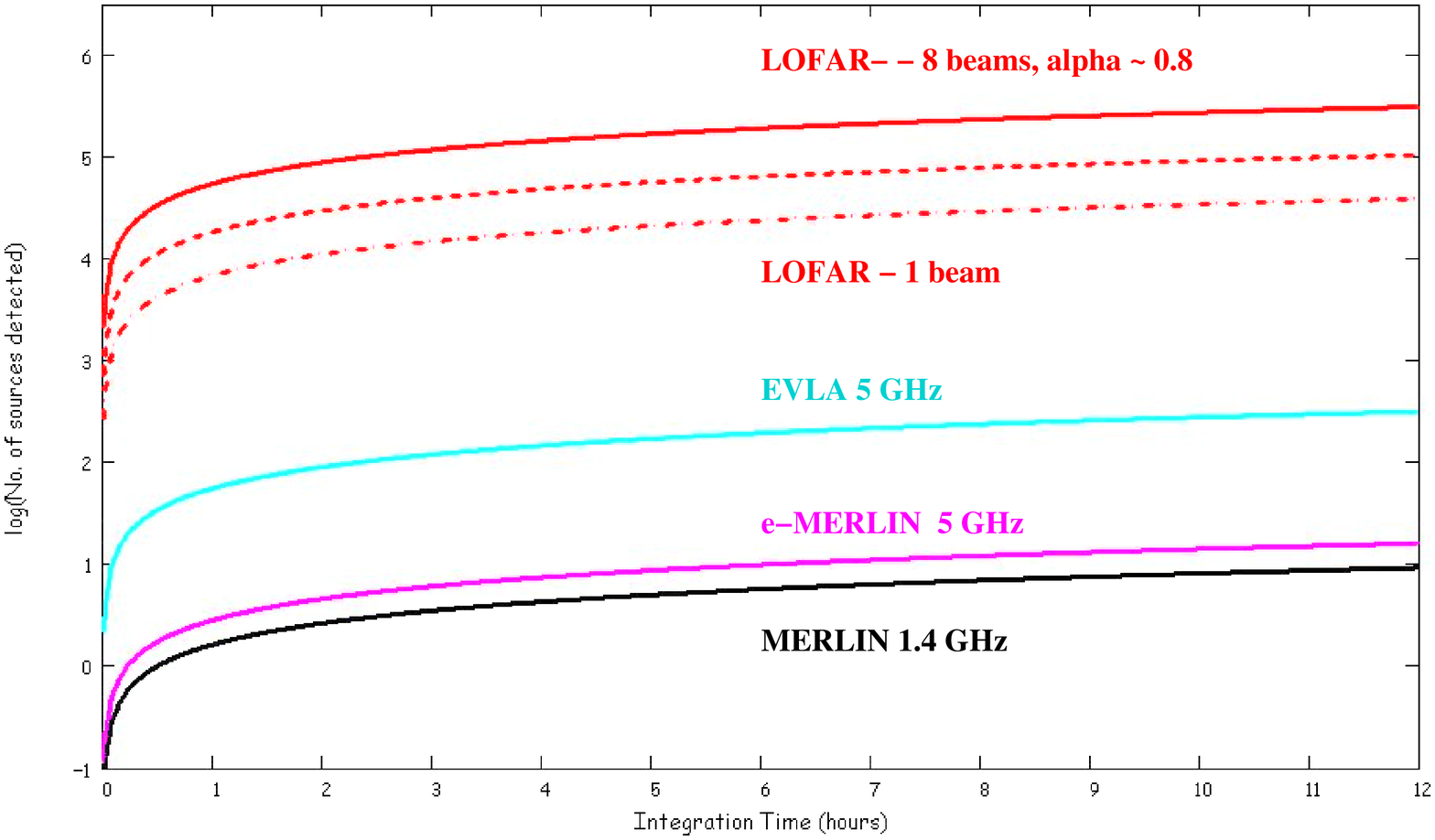}
      \caption{A plot of the number of sources in the field-of-view
   versus observing time (hours). LOFAR will detect in 1 hour as many
   star forming galaxies as current instruments (e.g. MERLIN) detect 
   only after many days of integration. Three curves are drawn for
   LOFAR (the top curve assumes a spectral index of -0.8 and includes
   contributions from 8 beams, the middle curves assumes a spectral
   index of -0.4, and the bottom curve represents the performances of a
   single beam.}
\label{fov}
   \end{figure}

Operating at the lowest radio frequencies, the LOFAR field-of-view
(FoV) is naturally much larger than either the EVLA or e-MERLIN. The
effect of this, together with multiple beams is shown in
Fig.~\ref{fov}. Assuming an effective dimension of 65 meters for each
LOFAR station, somewhere between $10^{4}$ and $10^{5}$ star forming
galaxies can be detected using all 8 beams. Again the numbers are
dependent on the assumed radio SED at cm wavelengths (see
Fig.~\ref{fov}). For the EVLA and e-MERLIN, the number of star forming
galaxies detected in the field of view is set to increase from a few
tens of sources per day to several hundred sources per day. The SKA
(with significantly more independently steerable beams) is likely to
have a similar capability to LOFAR, even at much higher frequencies.

%

\subsection{Confused ? You might very well be! } 
\label{confused} 

Source confusion usually provides a fundamental limit beyond which the
image noise level no longer improves - irrespective of the integration
times employed (Condon 1974). Confusion usually kicks in when the
surface density of sources (number of sources per beam area) exceeds
some limit, the exact figure depending on the slope of the source
counts. The steeper the count, the earlier confusion noise begins to
dominate the image (see Hogg 2001 and references therein).

\begin{figure}[b]
   \centering
   \vspace{6.3cm}
   \includegraphics{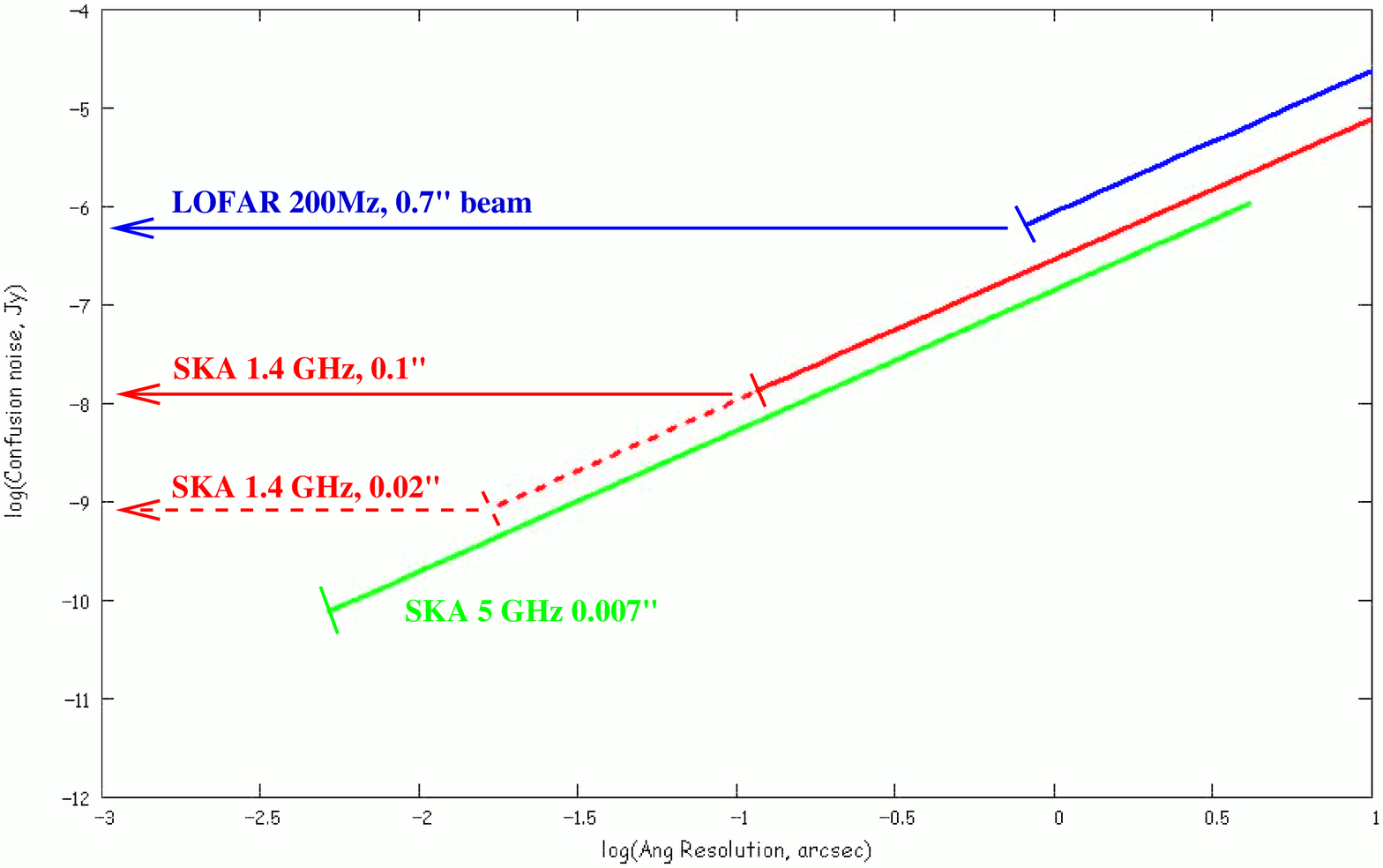}
      \caption{A plot of the $1\sigma$ confusion noise versus angular
        resolution for 200 MHz, 1.4 GHz and 5 GHz. After around 1000
        hours LOFAR is confusion limited (assuming a baseline length of
        400km). The SKA at 1.4 GHz reaches the confusion limit with the
        same baseline length after $\sim 1$ day. Assuming much longer
        integration times are employed in deep field continuum surveys,
        an angular resolution of 0.02 arcsec is required, implying a
        baseline length in excess of 1000 km.}
\label{confus}
   \end{figure}

It is important to realise that at these microJy levels of sensitivity
the radio sky literally lights-up. For example, in an observing run
with the $e$EVN of only 1 hour, there are potentially more than 1000
($5\sigma$) radio sources within the primary beam of the Effelsberg
100-m telescope. Fig.~\ref{confus} plots the confusion noise at 200
MHz, 1.4 and 5~GHz as a function of angular resolution. Again the SED
assumed for LOFAR is an important factor in determining how quickly the
confusion limit is reached. Here we have taken possibly the worst case
- $\alpha \sim -0.8$. Observing at 200~MHz and with a resolution of
$\sim 0.7$ arcsecs, LOFAR reaches the confusion noise ($\sim
1$~microJy) in 1000 hours (see also Rottgering 2002 for a discussion
about confusion and plans for a deep LOFAR continuum survey).
Similarly, the current SKA configuration (see Ekers 2002) specifies a
resolution of at least 0.1 arcsec at 1.4~GHz (equivalent to a baseline
length of 400~km at 1.4~GHz). Such a specification suggests the SKA
will hit the confusion limit rather quickly, $\sim 24$~hours at
1.4~GHz. It seems likely that a capable multiple-beam instrument such
as the SKA will permit much longer integration times to be employed,
similar to the 1000 hour long observations envisaged for LOFAR deep
field surveys. The inescapable conclusion to be drawn from
Fig.~\ref{confus} is that the SKA will require a significant fraction
of the array to be distributed over baselines of up to several thousand
km in extent.  Longer baselines will inevitably lead to a sparse SKA
array but the uv-coverage will still be good since large fractional
bandwidths can be employed. Indeed, the incorporation of longer
baselines might alleviate image dynamic range limitations -- the
brightest and most troublesome confusing sources at the edge of the
beam being largely resolved on baselines longer than a few thousand km.

%

\subsection{The need for high angular resolution} 
\label{size} 

While confusion places initial lower limits on the minimum baseline
length of next generation interferometers, we also need to consider
what optimum resolution is required to resolve and image these distant
star forming galaxies (not to mention other extra-galactic radio
sources, such as AGN that make up at least 20\% of the faint microJy
radio source population). The deep MERLIN and VLA detections of
moderate redshift microJy radio sources in the HDF-N (Muxlow et al.
1999), suggest that the angular size distribution of the majority of
sources peaks at about 1 arcsec with some sources clearly showing
compact sub-structure on sub-arcsecond scales. If more distant star
forming systems present radio emission on similar scales as local
nuclear starburst galaxies such as Arp~220, linear resolutions much
better than 500 pc will be required just to begin to probe the resolved
structures of these sources. At $z\sim 2$ the 500 pc subtends an angle
of $\sim 60$ milliarcsecond.  This is very similar to the resolution of
e-MERLIN at 5GHz. Combined e-MERLIN and EVLA observations should have
sufficient resolution and surface brightness sensitivity to begin to
probe the radio structures of these sources in some detail. In
addition, much higher resolution observations with the $e$EVN will be
able to distinguish between AGN and starburst activity in the
population of high-z, dust obscured, optically faint systems that only
reveal themselves in the radio and sub-mm wavebands. The observations
made by these upgraded instruments will be crucial in determining the
optimum distribution of SKA collecting area. Simple confusion noise
arguements already argue for baselines on scales of at least $\sim
1000$~km (see section\ref{confused}). Accurate astrometry, reliable
identification and complementarity with new sub-mm, near infra-red and
the next generation of extremely large (optical) telescopes (ALMA,
NGST, OWL, CELT etc) also present a strong case for an extended SKA.
With 2 orders of magnitude better sensitivity than any of its
predecessors, it would be short-sighted indeed to limit the resolution
of SKA to that currently employed by existing, connected
interferometers. With a resolution of ten milliarcsecond, the SKA will be
able to detect the sea of individual SNe, SNR, HII regions and GRBs
that will be the dynamic radio loud signature of massive star formation
in the early Universe.  Resolving such structures will provide crucial
clues to understanding the process of galaxy formation, perhaps 
distinguishing between monolithic collapse and hierarchical
merging processes. But in the meantime, the challenge of the next few
years will be to design a SKA that can do all this and more, without
significantly compromising the high brightness sensitivity required for
other programmes, such as narrow-band HI observations of both the
nearby and distant Universe.

\section{The radio Universe at $z > 6$ - beyond the epoch of re-ionisation} 
\label{caveats} 

I have confined my discussion to radio emission at $ z < 6$. While the
the FIR-radio correlation is now known to apply to at least $z \sim 1$
(Garrett 2002), at some critical redshift the synchrotron emission from
relativistic electrons is expected to be significantly reduced. In
particular, as inverse Compton losses (via the CMB) scale as
$(1+z)^{4}$, these will begin to dominate over synchrotron losses
beyond $z \sim 6$ (Carilli, Gnedin \& Owen 2002). The synchrotron
spectrum will steepen considerably and will be quenched all together in
regions where particle injection or re-acceleration has ceased. Prompt
emission from SNe and GRBs should still be detected however. It is also
interesting to note that at $z\sim 20$, the peak in the FIR dust
emission will begin to shift through the sub-mm wave-bands into the high
frequency radio part of the spectrum (see Fig.\ref{sens_z}). The
thought that next generation radio instruments might bask in the
favourable k-correction that sub-mm instruments such as SCUBA currently
enjoy (at $z\sim 1-10$) is an exciting one. However, at these early
epochs, star formation is in its infancy, and the dust content may
therefore be quite low. If there is emission from dust, we can also
expect the CMB to work against us again, setting a minimum grain
temperature and shifting the peak in the dust emission towards higher
frequencies at higher redshifts.  But to end on a positive note, it is
worth noting that radio emission from non-relativistic electrons (good
old free-free emission) will not be affected by IC losses, and its
relatively flat spectral index will make it ``future proof'' in terms
of the k-correction (Blain 2002).  Free-free emission will thus begin
to dominate the total radio emission that might be emitted by sources
at the very highest redshifts ($z > 10$).  As Blain (2002) also notes,
it may be one of the best sign posts to early galaxy formation at
redshifts greater than 10.

\begin{acknowledgements} 

I would like to thank Michiel van Haarlem for several useful discussion 
regarding LOFAR, in particular the array's multiple-beam capabilities. 

\end{acknowledgements}

\end{document}